\documentclass[%
reprint,
superscriptaddress,
altaffilletter,
showpacs,
showkeys,
amsmath, amssymb, aps,prl, floatfix,
]{revtex4-1}

\usepackage{graphicx}
\usepackage{dcolumn}
\usepackage{bm}
\usepackage{comment}
\usepackage[usenames,dvipsnames]{color}
\usepackage[dvipsnames]{xcolor}
\usepackage[normalem]{ulem}
\usepackage[symbol]{footmisc}
\usepackage{xspace}
\usepackage{ulem}
\usepackage{hyperref}
\hypersetup{colorlinks = true, allcolors = blue}
\usepackage[mathlines]{lineno}
\usepackage{multirow}
\DeclareGraphicsExtensions{.pdf,.png,.jpg}
\newcommand{\cupid}{\ensuremath{\rm CUPID\hbox{-}0}}

\newcommand{\exposure}{9.95~kg$\times$yr}

\newcommand{\startDAQ}{June 2017} 
\newcommand{\stopDAQ}{December 2018} 
\newcommand{\mvveff}{0.0763~^{+~0.0008}_{-~0.0010}}
\newcommand{\HalfLife}{T_{1/2}^{2\nu} = [8.60 \pm 0.03 \textrm{(stat.)}~^{+0.19}_{-0.13} \textrm{(syst.)}] \times 10^{19}~\textrm{yr}}

\newcommand{\xSSD}{255/254}
\newcommand{\xCA}{360/254}
\newcommand{\nbin}{287}

\newcommand{\CountsExp}{$ 14830 \pm 122 $}
\newcommand{\CountsSSD}{$ 14972 \pm 57 $}
\newcommand{\CountsCA}{$ 14095 \pm 56 $}
\newcommand{\vvactivitySSD}{[$8.71 \pm 0.03$ (stat.)]~mBq}
\newcommand{\vvactivityHSD}{[$8.74 \pm 0.03$ (stat.)]~mBq}
\newcommand{\ZS}{Zn$^{82}$Se}
\newcommand{\se}{$^{82}$Se}

\newcommand{\onu}{$0\nu\beta \beta$}

\newcommand{\QBB}{$\mathrm{Q}_{\beta\beta}$}

\newcommand{\vv}{2$\nu\beta\beta$\xspace}

\newcommand{\mspecone}{$\mathcal{M}_1$\xspace}
\newcommand{\mspeconea}{$\mathcal{M}_{1\alpha}$\xspace}
\newcommand{\mspeconeb}{$\mathcal{M}_{1\beta/\gamma}$\xspace}
\newcommand{\mspectwo}{$\mathcal{M}_2$\xspace}

\newcommand{\summspectwo}{$\Sigma_2$\xspace}

\begin{document}

\title{Evidence of Single State Dominance in the Two-Neutrino Double Beta Decay of Se-82 with CUPID-0}

\newcommand{\sapienza}{\affiliation{Dipartimento di Fisica, Sapienza Universit\`a di Roma, P.le Aldo Moro 2, 00185, Rome, Italy}}
\newcommand{\infnroma}{\affiliation{INFN Sezione di Roma, P.le Aldo Moro 2, 00185, Rome, Italy}}
\newcommand{\lnl}{\affiliation{INFN  Laboratori Nazionali di Legnaro, I-35020 Legnaro (Pd) - Italy}}
\newcommand{\lngs}{\affiliation{INFN  Laboratori Nazionali del Gran Sasso, I-67010 Assergi (AQ) - Italy}}
\newcommand{\lbl}{\affiliation{Lawrence Berkeley National Laboratory , Berkeley, California 94720, USA}}
\newcommand{\infnge}{\affiliation{INFN  Sezione di Genova, I-16146 Genova - Italy}}
\newcommand{\unige}{\affiliation{Dipartimento di Fisica, Universit\`{a} di Genova, I-16146 Genova - Italy}}
\newcommand{\infnmib}{\affiliation{INFN  Sezione di Milano Bicocca, I-20126 Milano - Italy}}
\newcommand{\unimib}{\affiliation{Dipartimento di Fisica, Universit\`{a} di Milano - Bicocca, I-20126 Milano - Italy}}
\newcommand{\csnsm}{\affiliation{CSNSM, Univ. Paris-Sud, CNRS/IN2P3, Universit\'e Paris-Saclay, 91405 Orsay, France}}
\newcommand{\cea}{\affiliation{IRFU, CEA, Universit\'e Paris-Saclay, F-91191 Gif-sur-Yvette, France}}
\newcommand{\gssi}{\affiliation{Gran Sasso Science Institute, 67100, L'Aquila - Italy}}
\newcommand{\usc}{\affiliation{Department of Physics  and Astronomy, University of South Carolina, Columbia, SC 29208 - USA\label{USC}}}
\newcommand{\fier}{\affiliation{Finnish Institute for Educational Research, P.O.Box 35 FI-40014 University of  Jyv\"askyl\"a, Finland}}
\newcommand{\ctp}{\affiliation{Center for Theoretical Physics, Sloane Physics Laboratory, Yale University, New Haven, Connecticut 06520-8120, USA}}
\newcommand{\paMartinez}{\affiliation{Present Address: Fundaci\'on ARAID and Laboratorio de F\'isica Nuclear y Astropart\'iculas, Universidad de Zaragoza, C/ Pedro Cerbuna 12, 50009 Zaragoza, Spain}}
\newcommand{\paNagorny}{\affiliation{Present Address: Queen's University, Physics Department, K7L 3N6, Kingston (ON), Canada}}
\newcommand{\paZolotarova}{\affiliation{Present Address: CSNSM, Univ. Paris-Sud, CNRS/IN2P3, Universit\'e Paris-Saclay, 91405 Orsay, France}}
\author{O.~Azzolini}\lnl
\author{J.W.~Beeman}\lbl
\author{F.~Bellini}\sapienza\infnroma
\author{M.~Beretta}\unimib\infnmib
\author{M.~Biassoni}\infnmib
\author{C.~Brofferio}\unimib\infnmib
\author{C.~Bucci} \lngs
\author{S.~Capelli}\unimib\infnmib
\author{L.~Cardani}\infnroma
\author{P.~Carniti}\unimib\infnmib
\author{N.~Casali}\infnroma
\author{D.~Chiesa}\unimib\infnmib
\author{M.~Clemenza}\unimib\infnmib
\author{O.~Cremonesi}\unimib\infnmib
\author{A.~Cruciani}\infnroma
\author{I.~Dafinei}\infnroma
\author{S.~Di~Domizio}\unige\infnge
\author{F.~Ferroni}\gssi\infnroma
\author{L.~Gironi}\unimib\infnmib
\author{A.~Giuliani}\csnsm
\author{P.~Gorla} \lngs
\author{C.~Gotti}\unimib\infnmib
\author{G.~Keppel}\lnl
\author{J.~Kotila}\fier\ctp
\author{M.~Martinez}\paMartinez\sapienza\infnroma
\author{S.~Nagorny}\paNagorny\gssi\lngs
\author{M.~Nastasi}\unimib\infnmib
\author{S.~Nisi}\lngs
\author{C.~Nones}\cea
\author{D.~Orlandi}\lngs
\author{L.~Pagnanini}\email[Corresponding author: ]{lorenzo.pagnanini@mib.infn.it}\unimib\infnmib
\author{M.~Pallavicini}\unige\infnge
\author{L.~Pattavina} \lngs
\author{M.~Pavan}\unimib\infnmib
\author{G.~Pessina}\infnmib
\author{V.~Pettinacci}\infnroma
\author{S.~Pirro}\lngs
\author{S.~Pozzi}\unimib\infnmib
\author{E.~Previtali}\unimib\infnmib
\author{A.~Puiu}\unimib\infnmib
\author{C.~Rusconi}\lngs\usc
\author{K.~Sch\"affner}\gssi\lngs
\author{C.~Tomei}\infnroma
\author{M.~Vignati}\infnroma
\author{A.~Zolotarova}\paZolotarova\cea

\date{\today}

\begin{abstract}
We report on the measurement of the two-neutrino double-$\beta$ decay of $^{82}$Se performed for the first time with cryogenic calorimeters, in the framework of the \cupid~experiment.
With an exposure of \exposure~of \ZS, we determine the two-neutrino double-$\beta$ decay half-life of $^{82}$Se with an unprecedented precision level, $\HalfLife$.
The very high signal-to-background ratio, along with the detailed reconstruction of the background sources allowed us to identify the single state dominance as the underlying mechanism of such process, demonstrating that the higher state dominance hypothesis is disfavored at the level of 5.5 $\sigma$.
\end{abstract}

\maketitle

Double-$\beta$ decay is a second-order weak process which changes two neutrons of a nucleus to protons, emitting two electrons. According to the Standard Model (SM) of particle physics, such decay is allowed only with the emission of two anti-neutrinos in the final state \cite{GoeppertMayer}.
The resulting nuclear transition is referred to as {\it two-neutrino double beta decay} (\vv), $(A,Z) \rightarrow (A,Z+2) + 2e^-+2\overline{\nu}_e$.
It has been observed for eleven isotopes with a half-life ranging from 10$^{18}$ to 10$^{24}$ yr \cite{Saakyan:2013yna,Barabash:2015eza,KamLAND-Zen:2019imh,NEMO-3:2019gwo}, that makes it the rarest nuclear weak process experimentally detected.
Conversely, the neutrinoless version of the decay (\onu) \cite{Furry:1939qr} is a lepton-number-violating process expected in several extensions of the SM but still never observed. The discovery of \onu~would demonstrate the Majorana nature of the neutrino, and would have important consequences for fundamental physics \cite{DellOro:2016tmg}.

Even if the experimental efforts are mainly devoted to \onu~searches, precision measurements of half-life and spectral shape for the SM-allowed \vv channel are of pivotal importance. Together with single-beta decay\cite{Gysbers:2019uyb}, \vv provides a useful benchmark to evaluate the reliability of the nuclear model calculations. It is indeed well known that the result has a strong dependence on the adopted approximations, and a precise measurement of \vv spectral shape offers a strong model-independent indication to understand which is the most realistic approximation. Double $\beta$-decay rate can be written as the product of a model-independent phase space factor ($G^{2\nu}$), and an effective nuclear matrix element ($\mathcal{M}^{eff}_{2\nu}$). The process is modeled as a sequence of two virtual $\beta$-decays going through one or more states of the $(A,Z+1)$ intermediate nucleus. The number of levels of the intermediate nucleus that contribute to the total transition amplitude is one of the assumptions that lead to different results in nuclear calculations. 
For \vv transition from ground to ground state, the intermediate levels are bound to have J$^P$=1$^+$. The \vv is called single state dominated (SSD) if it is governed by the lowest 1$^+$ energy level, higher-state dominated (HSD) otherwise. In the latter case the calculation is usually simplified by summing over all the virtual intermediate states and assuming an average closure energy \cite{Moreno:2008dz,Domin:2004za,Barea:2013bz}. 

Presently nuclear theories do not allow to establish with certainty whether a decay is SSD or HSD, thought for example in reference~\cite{Moreno:2008dz} arguments are given to discourage the SSD hypotheses for \se~and support it for $^{96}$Zr, $^{100}$Mo, and $^{116}$Cd. A powerful way to experimentally disentangle the two hypotheses is to study the energy distribution of the emitted electrons, that is slightly different in the two cases. The NEMO-3 collaboration \cite{Arnold:2004xq} has recently reported a clear indication for SSD in $^{100}$Mo \cite{NEMO-3:2019gwo}, while for \se~the result is affected by the limited statistics that prevents a clear discovery statement on the nuclear process details \cite{Arnold:2018tmo}.

In this letter, we report the measurement of the \se~\vv half-life performed by CUPID-0, an array of enriched scintillating ZnSe crystals operated as cryogenic calorimeters (or bolometers). These devices are single-particle detectors with a dual read-out. They combine the excellent energy resolution of bolometers, based on the thermal signal, with a particle identification capability, based on the shape of the scintillation signal \cite{Pirro:2005ar,Arnaboldi:2010jx,Beeman:2013vda,Artusa:2016maw,Armengaud:2017hit}.
Scintillating bolometers have been studied and optimized in view of an upgrade of the CUORE experiment \cite{Artusa:2014lgv}. CUORE is successfully operating a tonne-scale array of bolometers at 10~mK, proving that large size bolometric detectors are feasible and paving the way to a further improvement. CUORE Upgrade with Particle IDentification (CUPID) \cite{CUPIDInterestGroup:2019inu} will overcome the \onu~sensitivity limit of CUORE, due to $\alpha$-particles induced background, employing scintillating bolometers.

CUPID-0 is the first large scale demonstrator of CUPID. The detector is an array of 24 Zn$^{82}$Se crystals (95 $\pm$ 1$)\%$ enriched in $^{82}$Se and 2 ZnSe crystals with natural selenium, for a total mass of 10.5 kg. 
The ZnSe crystals are held in a copper frame through small PTFE clamps and laterally surrounded by Vikuiti$^{TM}$ plastic reflective foils. Germanium wafers , working as calorimetric light detectors \cite{Beeman:2013zva}, are interleaved with the ZnSe crystals.
Each calorimeter (Ge and ZnSe) is equipped with a Neutron Transmutation Doped (NTD) Ge thermistor \cite{Haller}, acting as temperature-voltage transducer.
The detector is suspended to the mixing chamber of an Oxford 1000 $^3$He/$^4$He cryostat operating at a base temperature of about 10 mK, and located underground in the Hall A of the Laboratori Nazionali del Gran Sasso (Italy). Details about the \cupid~detector can be found in Refs.~\cite{Arnaboldi:2017aek,Dafinei:2017xpc,Azzolini:2018tum,DiDomizio:2018ldc,Azzolini:2018dyb}.

The data we present here were collected between \startDAQ~and \stopDAQ~with an active mass of 8.74 kg of~\ZS, achieving \exposure~of \ZS~exposure. The total number of monitored \se~nuclei is (3.41~$\pm$~0.03)~$\times~10^{25}$. 
Each bolometer has an independent bias and read-out system \cite{Arnaboldi:2017aek}.
When a heat signal is triggered on a ZnSe crystal, the output of the correspondent light detector is recorded. These raw data are processed to obtain an energy calibrated spectrum as described in Refs.~\cite{Azzolini:2018yye,Azzolini:2018oph}.
The pulse height and shape parameters are estimated applying the matched-filter algorithm \cite{Gatti:1986cw}.
We establish the energy scale by fitting with a zero intercept parabolic function the positions of the most prominent gamma-lines registered in the range (511-2615) keV after exposing the detector to an external $^{232}$Th source. The uncertainty on the energy scale is investigated exploiting a dedicated calibration with a $^{56}$Co source. The distribution of residuals shows a parabolic dependence on energy, with a maximum deviation from zero of 3 keV between 511 keV and the \vv-decay endpoint at 3~MeV \cite{Azzolini:2019tta}. We take into account this effect by adding to the nominal energy of each event the corresponding residual, evaluated from the distribution. We tag time-coincident events simultaneously triggering different detectors within a 20 ms time window, previously optimized by studying the time distribution of double-hits physical events during $^{232}$Th calibrations. Time-coincident events are arranged in multiplets and used to build different spectra according to the number of events in their multiplet. In particular, the \mspecone spectrum includes the single-hit events, while \mspectwo and \summspectwo spectra are built with the double-hit events (\mspectwo comprises the energies detected by each crystal, \summspectwo the total energy released in two crystals). Finally, we tag the $\alpha$-particles relying on the light pulse shape parameter, defined in \cite{Azzolini:2018yye}, which provides a very effective $\alpha$-identification for particles with an energy greater than 2 MeV. To exploit this information, we split \mspecone data in two sub-spectra: \mspeconea, containing only $\alpha$-events with an energy $E > 2$~MeV, and \mspeconeb, filled with all the other single-hit events not identified as $\alpha$s.
We implement a series of data selection cuts in order to maximize our sensitivity to physics events \cite{Azzolini:2018yye}. The combined efficiency is constant above 150 keV and equal to $\varepsilon_{C}=(95.7 \pm 0.5)$ \%\cite{Azzolini:2019nmi}.

In order to measure the \vv activity, we perform a Bayesian fit \cite{Azzolini:2019nmi} to the experimental data with a linear combination of simulated spectra, which correspond to the ones produced in the detector by the \vv decay and by the background sources. The spectra normalization coefficients, determined through the fit, are the activities of the sources. The spectrum of each source is produced with a Geant4-based code which generates and propagates the particles in the \cupid~geometry until they are detected in the ZnSe crystals. We process the output of the Monte Carlo (MC) simulations in order to reproduce the specific features of the \cupid~detector (i.e.~pile-up, time coincidences, energy resolution, threshold, $\alpha$-identification, and efficiency). We exploit the detector modularity, the time correlation among events, and the particle identification capability to identify the expected signatures of the background sources. In particular, the contaminations of crystals produce distinctive peaks in the $\mathcal{M}_{1\alpha}$ spectrum, that constrains their activities. 
Moreover, we analyze the $\gamma$~lines to identify specific radioisotopes in the experimental setup. The $\mathcal{M}_{2}$ and $\Sigma_{2}$ spectra, in which the contribution from \vv decay is negligible, are fundamental to determine the cryostat and shields contaminations, producing double-hit events via Compton scattering or pair production. The muon contribution is normalized on the number of shower events that simultaneously trigger more than three crystals. The result of such normalization is compatible with the one obtainable using the muon flux at LNGS \cite{Ambrosio:1995cx}. 
We include the information from independent measurements by means of specific {\it Priors} \cite{Beeman:2015xjv,Azzolini:2018tum,Alduino:2016vtd}, and set non-negative uniform priors for the sources with unknown activity.
More details about the background model construction and the list of sources used to fit the CUPID-0 data are in Ref.~\cite{Azzolini:2019nmi}. The possible presence of pure $\beta$-emitters has been taken into consideration. The sources of this type are many and, producing a continuous spectrum without any distinctive signature, are all substantially degenerate for the purposes of our fit. Most of the pure $\beta$-emitters with a reasonably long half-life ($>$ 100 days), have a Q-value $< 700$ keV. In order to avoid source misidentification, we performed the fit of the \mspeconeb spectrum setting an energy threshold above 700 keV, so excluding most of the pure beta-emitters.
The only remaining pure $\beta$-emitter to be investigated is $^{90}$Sr, a fission product with 28.8 yr half-life that produces two consecutive $\beta$-decays $^{90}Sr\rightarrow^{90}Y\rightarrow^{90}$Zr with Q-values of 546~keV and 2281~keV respectively. Since there is no indication of fission products in our data, we consider a possible contamination of $^{90}$Sr as systematic uncertainty.

The fit interval ranges from 700 keV to 5 MeV for \mspeconeb, and from 2 MeV to 8 MeV for \mspeconea. \mspectwo and \summspectwo spectra include couples of coincident events with energies between 150 keV and 5 MeV. We adopt a variable step binning to include the counts of each $\gamma$- or $\alpha$-line in a single bin and to avoid bins with low counting statistics. As shown in the discussion of systematic uncertainties, the \vv activity measured by the fit is practically independent of those choices. The degrees of freedom of the global fit, which is performed simultaneously on the four spectra, correspond to the difference between the number of bins (\nbin) and the number of sources (33, i.e.~\se~\vv signal and 32 background sources). 

\begin{figure}[t!] 
\centering 
\includegraphics[width=0.48\textwidth]{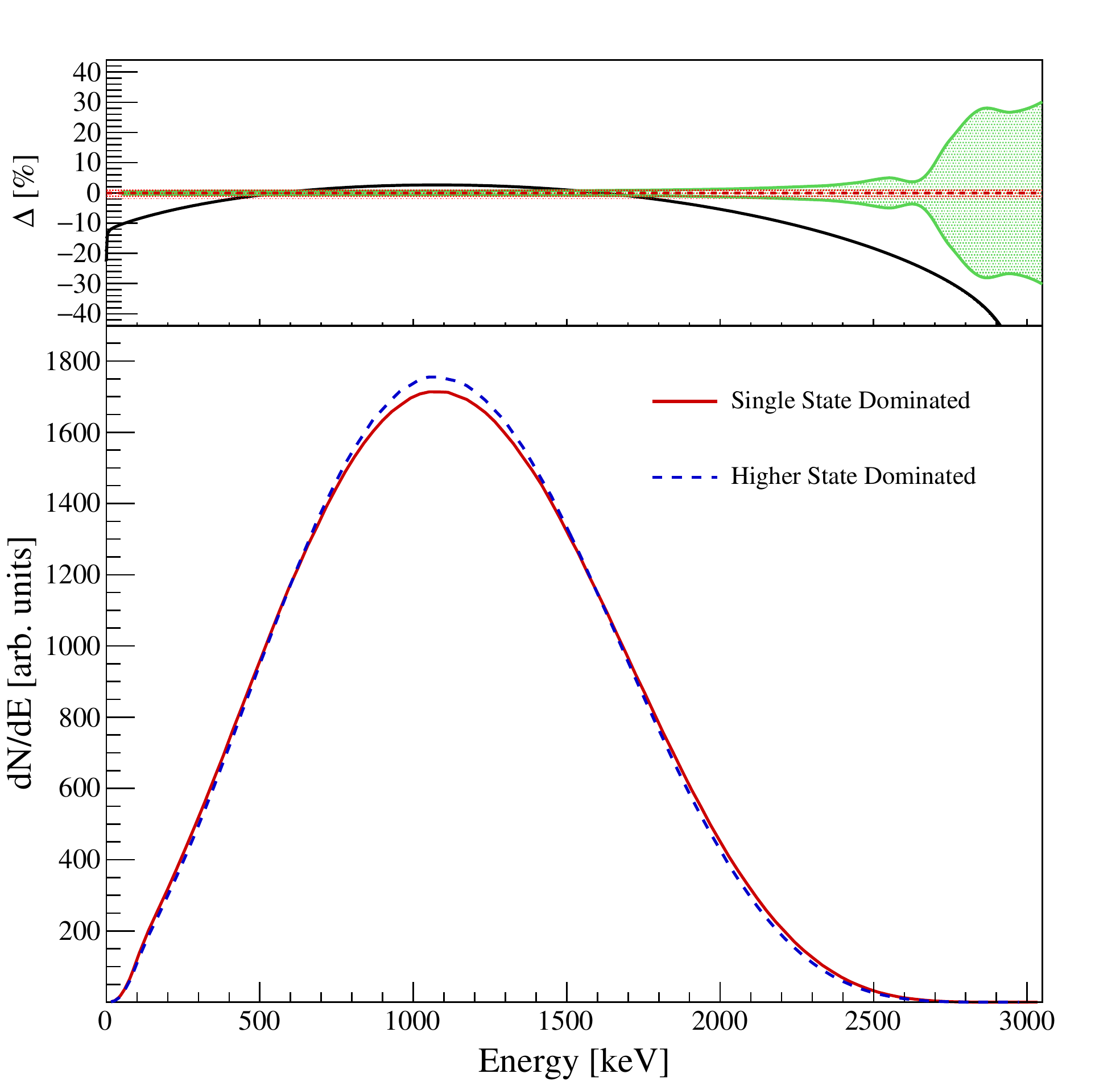}
\caption{The electrons sum spectra of \se~\vv simulated according to higher-state dominated (HSD, blue dashed line) and single state dominated (SSD, solid red line) hypothesis. In the top panel, we compare the percentage difference of HSD with respect to SSD spectrum (black solid line) with the statistical uncertainty of the collected data (green shaded band) and the total uncertainty on the \vv counts reconstructed in following analysis (red shaded band). The energy region from 2 MeV to the endpoint (\QBB) is the most sensitive to the different spectral shape.}
\label{fig:2nuModel}
\end{figure}

We simulate the \vv spectrum assuming the two different decay mechanisms, i.e.~SSD and HSD introduced above. We calculate both spectra using exact Dirac wave functions with finite nuclear size and electron screening as described in \cite{Kotila:2012zza}, exploiting the closure approximation (i.e.~an average higher state is chosen as the closure energy) in the HSD hypothesis. We compare the two \se~\vv spectra in Fig.~\ref{fig:2nuModel}, where they are reported as simulated in the \cupid~detector. In the top panel, the bin-to-bin ratio between the spectra shows that their percentage difference is more pronounced in the energy range from 2 MeV up to \QBB.  

\begin{figure}[!t] 
\centering 
\includegraphics[width=0.48\textwidth]{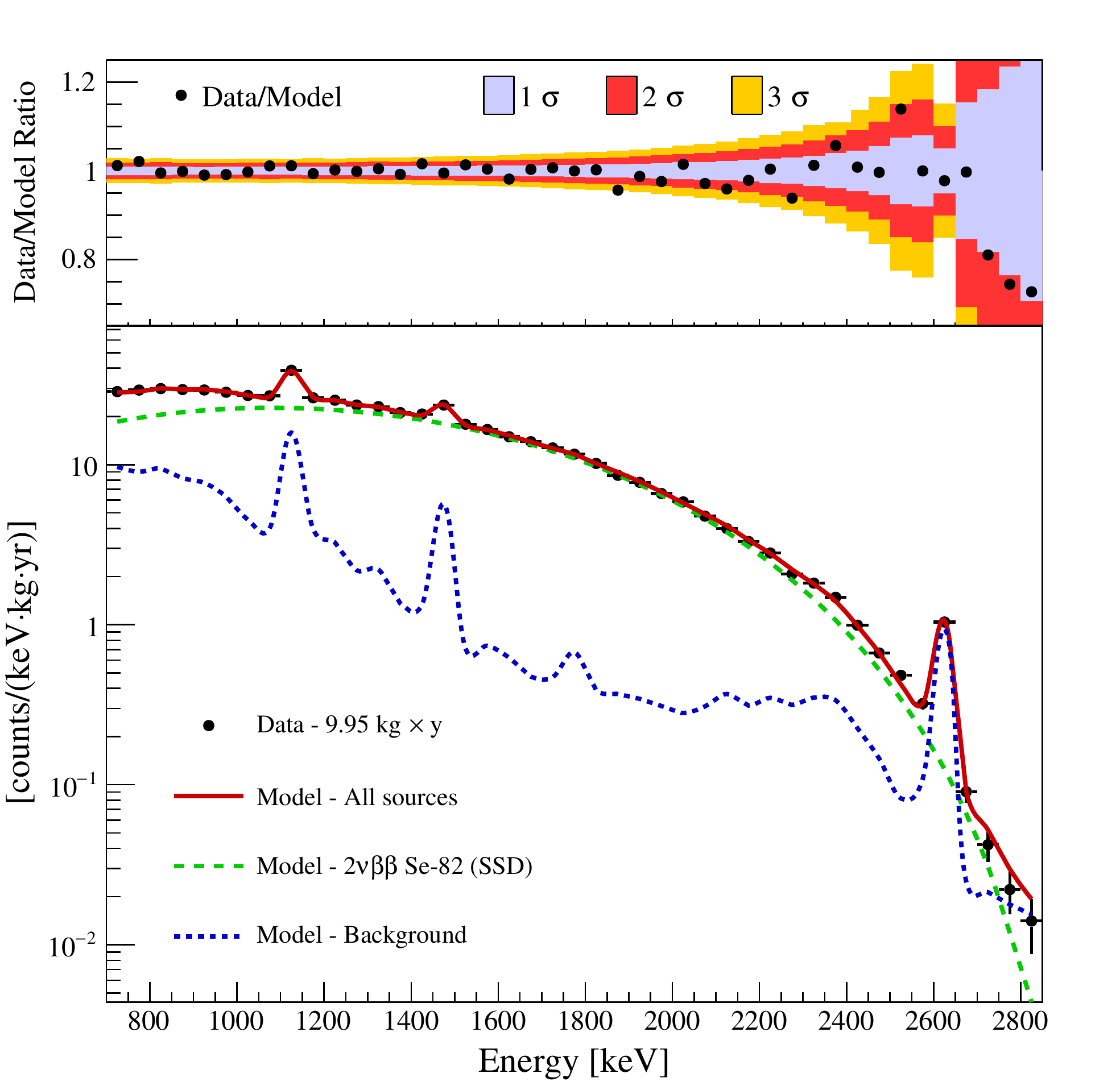}
\caption{Energy spectrum of the \mspeconeb events collected by \cupid~in \exposure~of \ZS~exposure (black dots). Only three $\gamma$-lines are clearly visible over the continuum due to \vv: $^{65}$Zn at 1116 keV, $^{40}$K at 1461 keV, and $^{208}$Tl at 2615 keV. The solid red line is the results of the Bayesian fit reconstruction with the SSD hypothesis for the \vv decay. The green line represents the \vv component, simulated assuming that the \vv is SSD. The blue line is the sum of the background sources. In the top panel, we show the bin-by-bin ratio between counts in the experimental spectrum and counts in the reconstructed one. The corresponding uncertainties at 1, 2, 3 $\sigma$ are shown as colored bands centered at 1.}
\label{fig:FitBM}
\end{figure}
\begin{figure}[t!] 
\centering 
\includegraphics[width=0.48\textwidth]{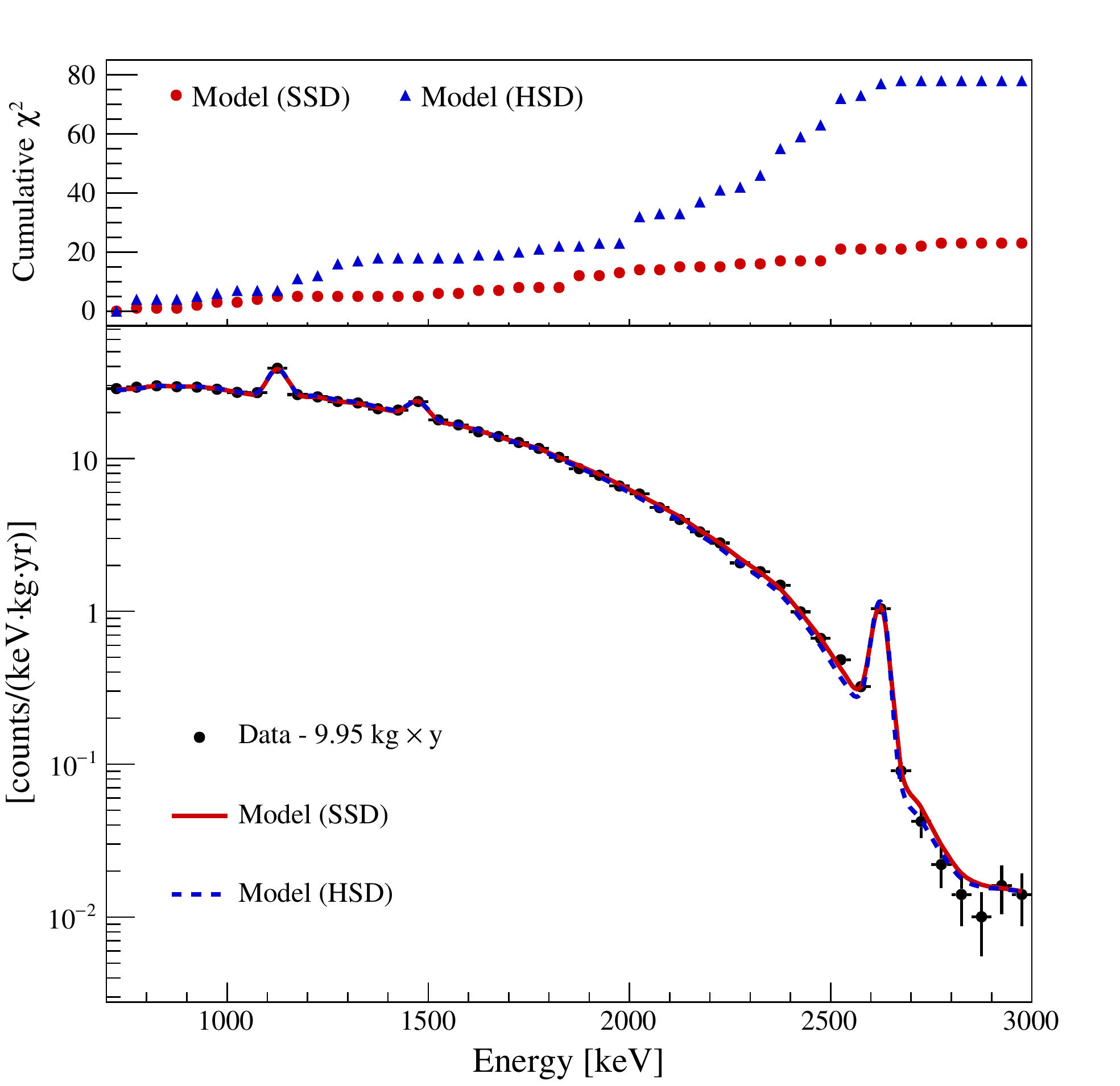}
\caption{Comparison between the \mspeconeb experimental spectrum (black dots) and the background model resulting from the fit, assuming the \vv is SSD (red line) or HSD (blue line), alternatively. In the top panel, we show the cumulative $\chi^{2}$ of the fits, calculated from 700 keV adding the pull squares of each bin to the ones of the previous bins. The $\chi^{2}$ vs.~Energy points out that SSD provides a much better description of the experimental data in the energy region above 2 MeV, where the difference between the models is more prominent.}
\label{fig:FitBM_HSD}
\end{figure}

We run the fit including the two \vv models alternatively.
In the SSD scenario, the fit reproduces the four experimental spectra very effectively with a global $\chi^{2}$/ndf = \xSSD~(calculated a posteriori as quality check). In particular, we show in Fig.~\ref{fig:FitBM} the excellent description of the spectral shape of \mspeconeb data, which drives the normalization of the \vv component. 
Conversely, the HSD does not provide a satisfactory description of the experimental data ($\chi^{2}$/ndf = \xCA). In Fig.~\ref{fig:FitBM_HSD} we compare the two reconstructions. As shown in the top panel the disagreement is more pronounced above 2 MeV, where we expect the largest difference between SSD and HSD models (Fig.~\ref{fig:2nuModel}). Considering the active mass of 8.74 kg of 95\% enriched \ZS, we measure a \vv activity of \vvactivitySSD~(SSD) and \vvactivityHSD~(HSD). The statistical uncertainty also includes the effect of the correlation among the \vv and the other spectra, since the \vv posterior is marginalized over the nuisance parameters (i.e.~the activities of the background sources).

We investigate the systematic uncertainty affecting the \vv activity performing the following fits in which:
\begin{itemize}
\item we simulate the contaminants in different positions of the cryostat and its shields;
\item we remove the sources resulting with an activity compatible with zero;
\item we include the $^{90}$Sr/$^{90}$Y contamination of ZnSe in the source list;
\item we use a fixed step binning for the \mspeconeb spectrum (15 - 50 keV);
\item we vary the threshold of the \mspeconeb spectrum (300, 400, 500, 600, 800, 900, 1000 keV);
\item we do not apply the $\alpha$-identification, thus fitting a unique $\mathcal{M}_{1}$ spectrum from 700 keV to 8 MeV;
\item we do not apply the energy scale correction; 
\item we use non-negative uniform priors for all the sources;
\end{itemize}

For each class of systematic effect, we quote the corresponding uncertainty as the maximum variation of the \vv activity with respect to the reference value. We also verified that the \vv activity evaluation is stable when fitting subsets of data. Particularly, by dividing the data in two halves corresponding to the first and second part of data taking, or selecting different group of detectors, we obtain results fully compatible from the statistical point of view. We evaluate the combined systematic uncertainty of the fit adding in quadrature all the uncertainties listed in Tab.~\ref{tab:syst}. 
Finally, we include the uncertainties on theoretical description of the \vv decay (1.0 \% \cite{Doi:1985dx}), efficiency calculation and \se~nuclei, added in quadrature.

To investigate the compatibility of the two models with the data, we compare the experimental counts ($N_{exp}$) in the range between 2 and 3 MeV with the ones predicted by the two models ($N_{X}$ where $X=SSD$ or $HSD$). We quantify the accordance between data and model through the parameter:
\begin{equation}
t_X = \frac{|N_{exp}-N_{X}|}{\sqrt{\sigma^{2}_{exp}+\sigma^{2}_{X}}},
\end{equation}
where $\sigma_{exp} = \sqrt{N_{exp}}$, and $\sigma_{X}$ is the statistical uncertainty of the counts predicted by the model.

In the different fits performed to quantify the systematic effect (Tab.~\ref{tab:syst}), $t_{HSD}$ spans from 6.6 to 5.5, while $t_{SSD}$ is always of the order of 1. The results obtained from the fit configuration that returns the lowest value for HSD are reported in Tab.~\ref{tab:ModelCounts}. 
To investigate the sensitivity of the experiment to reject the HSD hypothesis, we performed a toy MC in which 10$^{6}$ experiments have been simulated by generating Poisson distributed experimental counts in the [2 -- 3] MeV range. For each simulated experiment, we computed the value of $t_{HSD}$, taking into account the statistical and systematic uncertainties of the model reconstruction. The central value of resulting distribution of the t-parameter is 6.1, and the probability to obtain $t>3$ is $>99.8\%$. The very high statistical significance of this result, allows to discover that the \vv of \se~is single state dominated, ruling out the HSD hypothesis.
We convert the experimental value of \vv activity and its uncertainty in \se~\vv half-life,
\begin{equation}
\HalfLife.
\end{equation}
This value is compatible at 1.3 $\sigma$ with the NEMO-3 result \cite{Arnold:2018tmo}, but the statistical and systematic uncertainties are improved by a factor of six and three, respectively. Finally, we determine $\mathcal{M}^{eff}_{2\nu}$ for the \vv of \se~to be $\mathcal{M}^{eff}_{2\nu} = \mvveff$, calculating $G^{2\nu} = (1.996 \pm 0.028)\times10^{-18}$ y$^{-1}$ under the SSD model \cite{Kotila:2012zza}.

Since the \vv of \se~has been not expected to be single-state dominated, the $\mathcal{M}^{eff}_{2\nu}$ calculated up to now in the framework of different models (IBM \cite{Barea:2015kwa}, ISM \cite{Caurier:2011gi,Menendez:2009xa} and QRPA\cite{Simkovic:2013qiy,Simkovic:2018hiq}) can not be compared with our result. This will be a useful benchmark for the nuclear models, when $\mathcal{M}^{eff}_{2\nu}$ calculations will be available. 

\begin{table}
\begin{center}
\caption{Systematic uncertainties affecting the \vv activity measurement due to fit parameterization. For each class of test, we calculate the maximum deviation of the \vv activity with respect to the reference value. We obtain the combined value by summing in quadrature the results of each class. We also quote the uncertainty on the selection efficiency and the number of \se~nuclei. In the last row, we quote the total systematic uncertainty, given by summing in quadrature the listed contributions.}
\begin{tabular}{llr}
\hline
\hline
&Systematic Source & $\Delta A_{2\nu}$ \\
\noalign{\smallskip}\hline\noalign{\smallskip}
{\bf Fit}& Source localization & $^{+0.36}_{-0.21}$ \%\\
\noalign{\smallskip}
&Reduced sources list & $-$0.10 \%\\
\noalign{\smallskip}
& $^{90}$Sr/$^{90}$Y & $-$1.57 \%\\
\noalign{\smallskip}
&Fixed step binning & $+$0.16 \%\\
\noalign{\smallskip}
&Threshold of \mspeconeb & $+$0.15 \%\\
\noalign{\smallskip}
&$\alpha$-identification & $-$0.01 \%\\
\noalign{\smallskip}
& Energy scale & $-$0.39 \%\\
\noalign{\smallskip}
& Prior distributions & $+$0.04 \%\\
\noalign{\smallskip}\cline{2-3}\noalign{\smallskip}
& Combined & $^{+0.4}_{-1.6}$ \%\\
\noalign{\smallskip}\hline\noalign{\smallskip}
{\bf Detector}&Efficiency & $\pm$0.5 \%\\
&$^{82}$Se atoms & $\pm$1.0 \%\\
\noalign{\smallskip}\hline\noalign{\smallskip}
{\bf Model}& \vv  & $\pm 1.0$ \%\\
\noalign{\smallskip}\hline\noalign{\smallskip}
{\bf Total} & &  $^{+1.6}_{-2.2}$ \%\\
\noalign{\smallskip}\hline
\hline
\end{tabular}
\label{tab:syst}
\end{center}
\end{table}

\begin{table}[t!]
\begin{center}
\caption{Comparison between the experimental counts of the \mspecone~spectrum from 2 to 3 MeV and the expected ones by the model, assuming that \vv is SSD or HSD, alternatively. We report only the results of the fit in which we choose a threshold of 900 keV, since this returns the lowest value of the $t$-parameter for HSD.}
\begin{tabular}{ccc}
\hline
\hline
Spectrum &  Counts & $t$[$\sigma$]\\
\noalign{\smallskip}\hline
{\bf Experimental} &  \CountsExp &  \\
\noalign{\smallskip}
{\bf Model (SSD)}  &  \CountsSSD & 1.1\\
\noalign{\smallskip}
{\bf Model (HSD)}  &  \CountsCA  & 5.5\\
\noalign{\smallskip}\hline
\hline
\end{tabular}
\label{tab:ModelCounts}
\end{center}
\end{table}
In summary, we have performed the most precise measurement of the \se~\vv half-life, with an uncertainty of 2.2\%. Such precision level is the best ever obtained among the \vv measurements of $^{76}$Ge by GERDA (4.9\% \cite{Agostini:2015nwa}), $^{100}$Mo by LUMINEU (5.8\% \cite{Armengaud:2017hit}), $^{130}$Te by CUORE-0 (7.7\% \cite{Alduino:2016vtd}), $^{136}$Xe by EXO-200 (2.8\% \cite{Albert:2013gpz})).
Moreover, we have established that the \vv of \se~is single state dominated, ruling out the hypothesis that the higher states of the intermediate nucleus participate to this nuclear transition.
Such results are based on a solid model of the CUPID-0 background \cite{Azzolini:2019nmi} and are achieved operating ultra-pure scintillating cryogenic calorimeters, highly enriched in \se, with detailed control of the radioactive contaminations of the materials. 
The wide span of physics results obtained, despite the small exposure, proves once more the potential of cryogenic calorimeters, setting an important milestone for the next-generation CUPID experiment.

\begin{acknowledgments}
This work was partially supported by the European Research Council (FP7/2007-2013) under contract LUCIFER no. 247115. The work of JK was supported by the Academy of Finland (Grant No. 314733). We thank Professor F. Iachello for suggesting the investigation of the SSD model, M. Iannone for his help in all the stages of the detector assembly, A. Pelosi for constructing the assembly line, M. Guetti for the assistance in the cryogenic operations, R. Gaigher for the mechanics of the calibration system, M. Lindozzi for the cryostat monitoring system, M. Perego for his invaluable help in many tasks, the mechanical workshop of LNGS (E. Tatananni, A. Rotilio, A. Corsi, and B. Romualdi) for the continuous help in the overall set-up design. We acknowledge the Dark Side Collaboration for the use of the low-radon clean room. This work makes use of the DIANA data analysis and APOLLO data acquisition software which has been developed by the CUORICINO, CUORE, LUCIFER and, \cupid~collaborations. This work makes use of the Arby software for Geant4 based Monte Carlo simulations, that has been developed in the framework of the Milano – Bicocca R\&D activities and that is maintained by O. Cremonesi and S. Pozzi.
\end{acknowledgments}
\bibliography{main}
\end{document}